\documentclass[useAMS]{mn2e}
\usepackage{graphicx}
\usepackage{amsmath}

\title[Suzaku view of GX~304-1]
{{\it Suzaku} view of Be/X-ray binary pulsar GX~304-1 during Type~I X-ray outbursts}
\author[G. K. Jaisawal, S. Naik and P. Epili]{Gaurava K. Jaisawal\thanks{gaurava@prl.res.in}, Sachindra Naik\thanks{snaik@prl.res.in} and Prahlad Epili\thanks{prahlad@prl.res.in} \\
Astronomy and Astrophysics Division, Physical Research Laboratory, Navrangapura, Ahmedabad - 380009, Gujarat, India\\}

\begin{document}

\date{}

\maketitle

\begin{abstract}

We report the timing and spectral properties of Be/X-ray binary pulsar 
GX~304-1 by using two {\it Suzaku} observations during its 2010 August and 
2012 January X-ray outbursts. Pulsations at $\sim$275~s were clearly 
detected in the light curves from both the observations. Pulse profiles 
were found to be strongly energy-dependent. During 2010 observation, 
prominent dips seen in soft X-ray ($\leq$10 keV) pulse profiles were 
found to be absent at higher energies. However, during 2012 observation, 
the pulse profiles were complex due to the presence of several dips.
Significant changes in the shape of the pulse profiles were detected 
at high energies ($>$35 keV). A phase shift of $\sim$0.3 was detected
while comparing the phase of main dip in pulse profiles below and above
$\sim$35 keV. Broad-band energy spectrum of pulsar was well described by 
a partially absorbed Negative and Positive power-law with Exponential cutoff 
(NPEX) model with 6.4~keV iron line and a cyclotron absorption feature. Energy 
of cyclotron absorption line was found to be $\sim$53 and 50~keV for 2010 
and 2012 observations, respectively, indicating a marginal positive dependence 
on source luminosity. Based on the results obtained from phase-resolved spectroscopy,
the absorption dips in the pulse profiles can be interpreted as due to the
presence of additional matter at same phases. Observed positive correlation between 
cyclotron line energy and luminosity, and significant pulse-phase variation of cyclotron 
parameters are discussed in the perspective of theoretical models on cyclotron 
absorption line in X-ray pulsars.  

\end{abstract}


\begin{figure*}
\centering
\includegraphics[height=5in, width=4.8in, angle=-90]{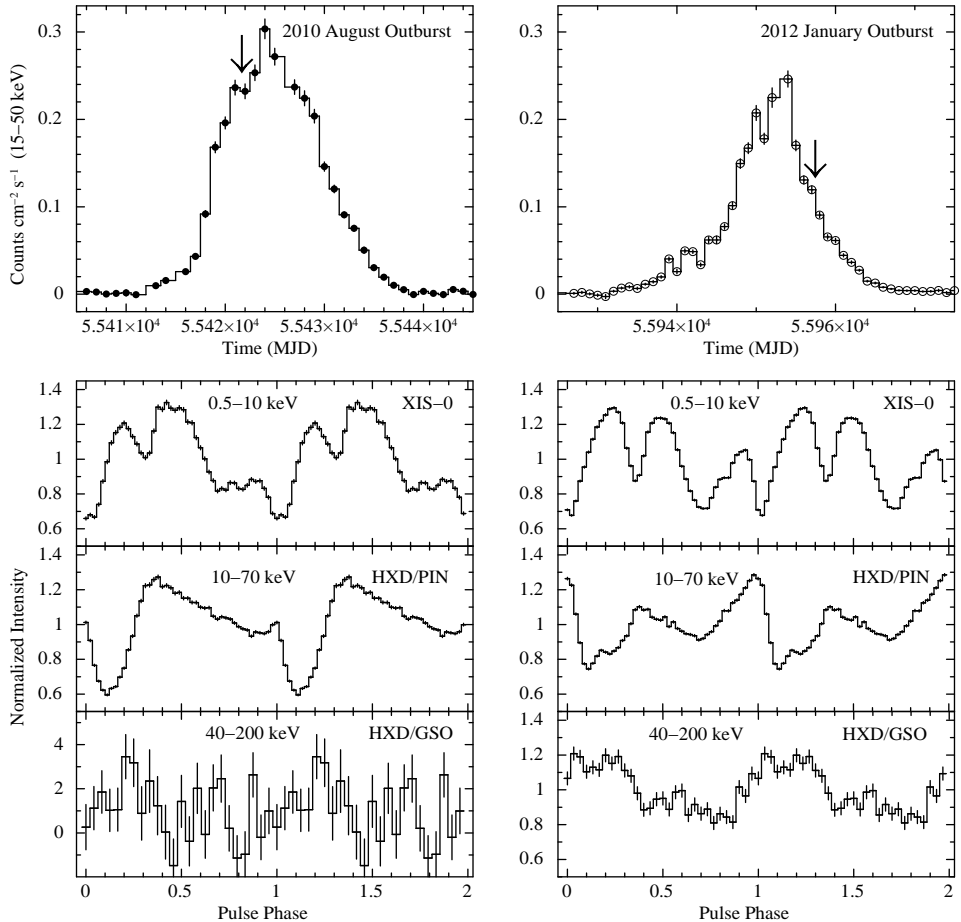}
\caption{$Swift/BAT$ light curves of GX~304-1 in the 15-50 keV energy band, 
from 2010 July 28 (MJD 55405) to 2010 September 06 (MJD 55445) and 2011 
December 30 (MJD 55925) to 2012 February 18 (MJD 55975) are shown in both 
sides of first panel, respectively. The arrow marks in both sides of first 
panel shows the date of {\it Suzaku} observations of GX~304-1 during outburst. 
Corresponding pulse profiles in 0.5-10~keV (XIS-0; second panel), 10-70~keV 
(PIN; third panel) and 40-200~keV (GSO; fourth panel) obtained from the 
background subtracted light curves of both the observations are shown in both 
sides of figure. Phase zero was arbitrarily chosen at MJD 55421.76 and 
55957.4326 for first and second observations, respectively. The errors in the 
pulse profiles are estimated for 1$\sigma$ confidence level and two pulses are 
shown for clarity.}
\label{lc}
\end{figure*}
\begin{figure*}
\centering
\includegraphics[height=6.5in, width=4.6in, angle=-90]{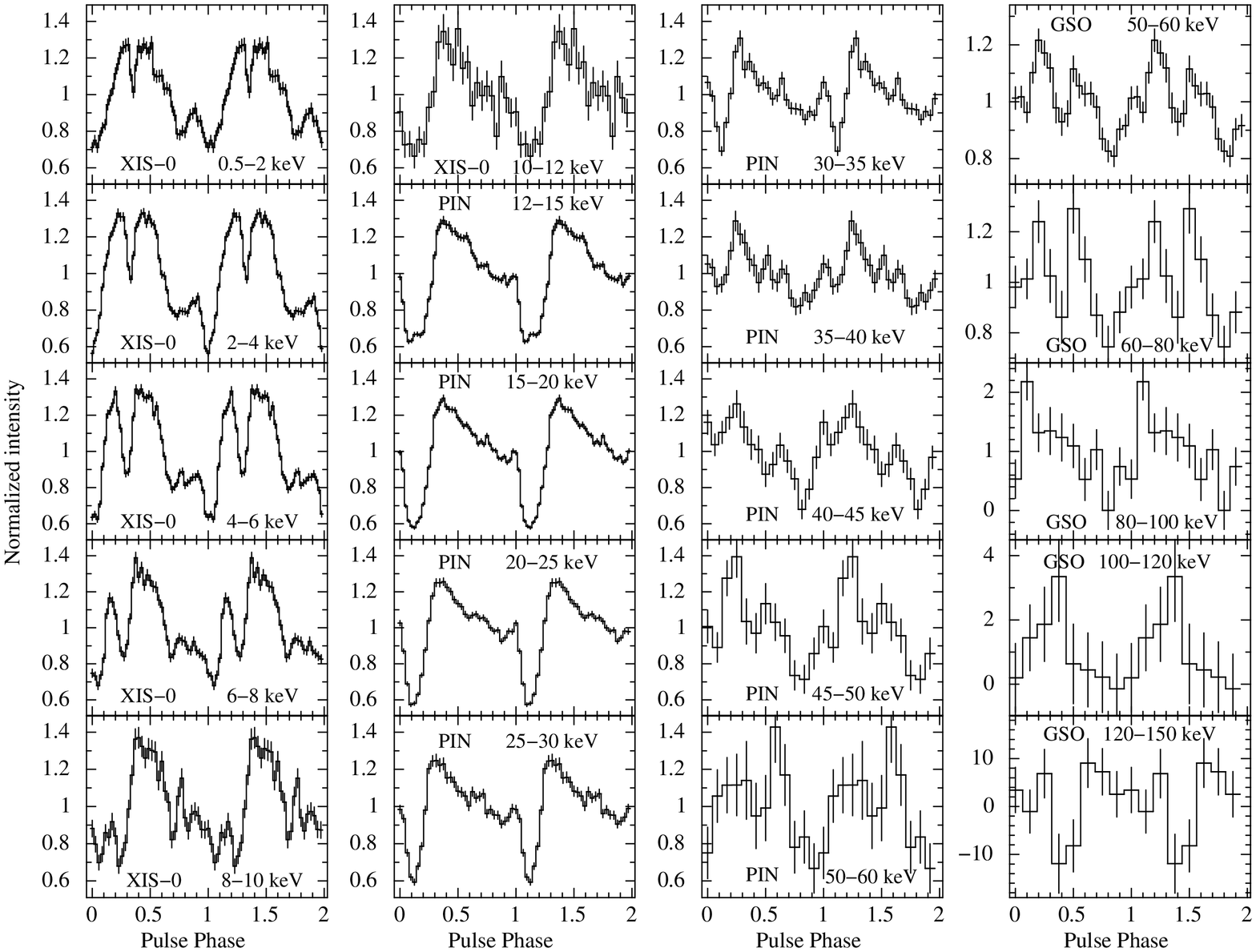}
\caption{Energy-resolved pulse profiles of GX~304-1 obtained from XIS-0, HXD/PIN and 
HXD/GSO light curves at various energy ranges, during first {\it Suzaku} observation in 
2010 August outburst. The presence of absorption dips in profiles can be seen at 
various pulse phases. The error bars represent 1$\sigma$ uncertainties. Two pulses 
in each panel are shown for clarity.}
\label{erpp2010}
\end{figure*}

\section{Introduction}

Be/X-ray binary pulsar GX~304-1 was discovered during hard X-ray sky surveys 
with balloon observations in 1967 (McClintock et al. 1971). The X-ray source 
was detected with successive space missions and recognized as 3U~1258-61, 
4U~1258-61 \& 2S~1258-613 (Giacconi et al. 1974; Bradt et al. 1977; 
Forman et al. 1978). Using data from $SAS-3$ observations, X-ray pulsations 
at $\sim$272~s were discovered in the source (McClintock et al. 1977). 
Spectral investigation of the pulsar, carried out from balloon observations, 
showed that the continuum spectrum in 18-35~keV range was described by a 
power-law (Maurer et al. 1982). Later, a power-law model modified with high 
energy cutoff was used to explain the 2-40~keV continuum spectrum obtained 
from  $HEAO~1$ observation of the pulsar (White, Swank \& Holt 1983). Analyzing 
the periodicity of the X-ray outbursts in 7 years of data from $Vela~5B$, the 
orbital period of the GX~304-1 was reported to be 132.5~d (Priedhorsky \& 
Terrell 1983). A shell star with visual magnitude of 15 was discovered in 
the X-ray error box of the neutron star and identified as the optical 
companion of the pulsar (Mason et al. 1978). High resolution optical 
spectroscopy established the spectral class of the companion as Be star 
of type B2~Vne which is at a distance of 2.4$\pm$0.5~kpc (Parkes et al. 
1980). 

GX~304-1 was monitored with $EXOSAT$ covering a duration of an expected outburst 
in 1984 July/August (Pietsch et al. 1986). However, no X-ray outburst was
detected during the $EXOSAT$ monitoring campaign. In contrast, the source 
flux was estimated to be significantly low e.g. by a factor of 25, than the 
quiescent flux level. This observed peculiarity was characterized as the X-ray 
``off'' state of the pulsar. Long term optical monitoring of the Be companion 
star in 1978-1988 suggested a major change (loss) in the Be envelop or circumstellar 
disk, the consequence of which is considered as the possible cause of the X-ray 
``off'' state in GX~304-1 (Corbet et al. 1986; Pietsch et al. 1986; Haefner 1988). 

After 28 years of quiescence, an X-ray outburst was detected from GX~304-1 with the
{\it INTEGRAL} observatory in 2008 June (Manousakis et al. 2008) after which the source 
was found to be active in X-rays. Since then, several X-ray outbursts have been detected 
in the pulsar with {\it Swift}/BAT and $MAXI$ (Yamamoto et al. 2009, 2012; Krimm et al. 
2010; Mihara et al. 2010). Using the {\it Rossi X-Ray Timing Explorer (RXTE)} observations 
during 2010 August outburst, energy and luminosity dependence of pulse profiles were found 
in GX~304-1 (Devasia et al. 2011). Apart from the evolution of pulse profiles, a 
quasi-periodic oscillation (QPO) at $\sim$0.125~Hz was detected with harmonics in 
several $RXTE$/PCA observations during this outburst. The pulsar spectrum in 3-30~keV 
range was described with a partial covering high energy cutoff power-law 
model (Devasia et al. 2011). During the same outburst in 2010 August, a cyclotron 
absorption feature at $\sim$54~keV was detected in the pulsar spectrum (Yamamoto et al. 
2011) and the corresponding magnetic field of the neutron star was estimated to be 
$\sim$4.7$\times$10$^{12}$~G. A positive correlation between cyclotron energy and 
luminosity was seen during 2012 January-February outburst with {\it INTEGRAL} 
(Klochkov et al. 2012). Malacaria et al. (2015) performed timing and 
spectral analysis of the pulsar using the same {\it INTEGRAL} data. The shape 
of the pulse profiles obtained from these observations were found to be similar 
in 20-40~keV, 40-60~keV and 18-80~keV energy ranges. Phase-resolved spectroscopy 
was carried out by stacking multiple spectra of different fluxes. From this analysis,
the cyclotron absorption line energy was found to be nearly constant (within errors) 
with pulse phases, except about 10\% variation at one phase bin (Malacaria et al. 
2015). The evolution of the pulse period with luminosity was studied by using 
data from $MAXI$/GSC, $RXTE$/PCA, $Swift$/XRT and $Fermi$/GBM observations 
during a series of outbursts from 2009 to 2013 (Postnov et al. 2015; Sugizaki 
et al. 2015). The observed pulse period variation was interpreted in terms of binary 
modulation along with the spinning-up of the neutron star. The orbital parameters of 
the binary system were estimated to be -- orbital period = 132.19~d, epoch at the 
periastron = MJD~55425, pulse period $\simeq$275.45~s, orbital eccentricity 
$\simeq$ 0.5, a$_x$ sin~$i$ $\simeq$ 500--600~light-s and  $\omega$ at periastron 
$\simeq$ 122.5$^\circ$--130$^\circ$ (Sugizaki et al. 2015). 

In this work, the timing and broad-band spectral properties of GX~304-1 were 
presented in detail by using two {\it Suzaku} observations during its outbursts in 
2010 August and 2012 January. Earlier, the evolution of pulse profiles up to 
$\sim$30~keV had been reported by using $RXTE$/PCA observations during 2010 
outburst (Devasia et al. 2011). Using {\it Suzaku} observations, the 
evolution of pulse profiles up to $\sim$150~keV has been presented in this paper.
The observed changes in the shape of pulse profiles close to the cyclotron absorption 
line are also being discussed. Although spectral studies of the pulsar during its 2012 
outburst has been reported in 5-100~keV range by using {\it INTEGRAL} observations 
(Malacaria et al. 2015), {\it Suzaku} observations provide a better opportunity to 
investigate a detailed spectral and timing study in broad energy range (1-150 keV). 
The details of observations, data analysis, results and interpretations are described 
in following sections of the paper.

\begin{figure*}
\centering
\includegraphics[height=6.5in, width=4.6in, angle=-90]{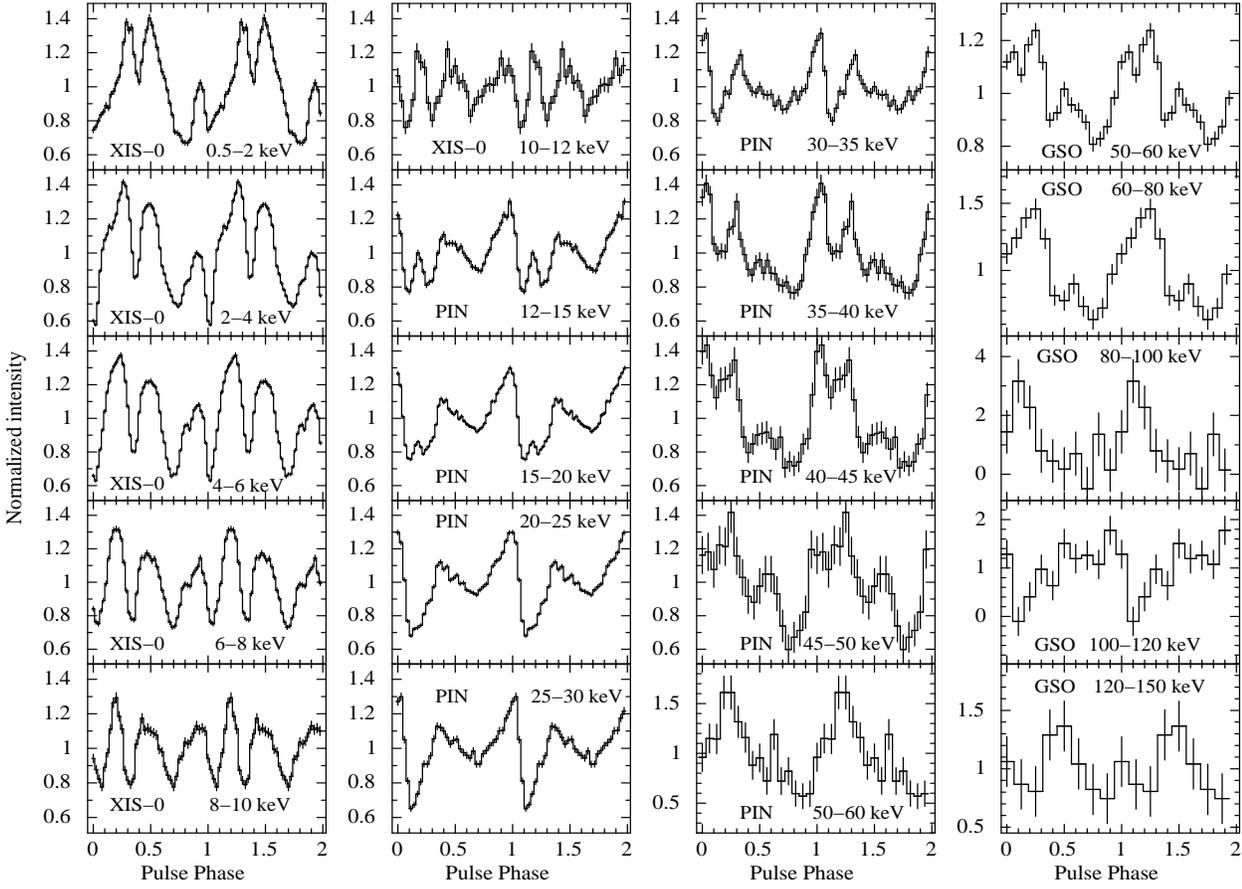}
\caption{Energy-resolved pulse profiles of GX~304-1 obtained from XIS-0, HXD/PIN 
and HXD/GSO light curves at various energy ranges, during second {\it Suzaku} observation 
in 2012 January outburst. The presence of absorption dips in profiles at higher energies 
can be seen at various pulse phases. The error bars represent 1$\sigma$ uncertainties. 
Two pulses in each panel are shown for clarity.}
\label{erpp2012}
\end{figure*}

\section{Observation and Analysis}

{\it Suzaku}, the fifth Japanese X-ray mission, was launched by Japan Aerospace 
Exploration Agency (JAXA) on 2005 July 10 (Mitsuda et al. 2007). It offers a 
broad energy coverage (0.2-600~keV) in the electromagnetic spectrum by using
two sets of instruments e.g. the X-ray Imaging Spectrometers (XIS; Koyama et 
al. 2007) and Hard X-ray Detectors (HXD; Takahashi et al. 2007). The XISs are 
imaging CCD cameras and cover 0.2-12~keV energy range. Among four XISs, three 
XISs (XIS-0, XIS-2, XIS-3) are front-illuminated whereas one XIS (XIS-1) is 
back-illuminated. The effective areas of front and back-illuminated XISs are 
340~cm$^2$ and 390~cm$^2$ at 1.5~keV, respectively. Field of view of XIS detectors 
is $18'\times18'$ in full window mode. The HXD unit of {\it Suzaku} consists of two 
sets of detectors such as HXD/PIN and HXD/GSO. HXD/PIN is a silicon diode detector 
covering 10-70~keV energy range, whereas HXD/GSO is a crystal scintillator detector 
working in 40-600~keV energy range.  Effective areas for HXD/PIN and HXD/GSO are 
145 cm$^2$ at 15 keV and 315 cm$^2$ at 100 keV, respectively. Field of view of 
HXD/PIN and HXD/GSO (up to 100~keV) is $34'\times34'$.

GX~304-1 was observed with {\it Suzaku} during its outbursts in 2010 August and 
2012 January. $Swift$/BAT light curves of the pulsar in 15-50~keV range, 
covering the outbursts are shown in top panels of Fig.~\ref{lc}.
Arrow marks in both panels indicate the date of {\it Suzaku} observation 
of GX~304-1 during respective outbursts. During 2010 August outburst, the 
{\it Suzaku} observation was carried out at the peak of outburst whereas the
second observation was made during the decay phase of the 2012 January 
outburst. The first observation was performed on 2010 August 13, in 
`HXD nominal' position with an effective exposure of $\sim$5.1~ks and 
$\sim$12.9~ks for XIS and HXD detectors, respectively. However, the second
observation was made during 2012 January 31-February 02 for a longer effective 
exposures of $\sim$16.5~ks for XIS and $\sim$58.7~ks for HXD. The 
second observation was performed in `XIS nominal' position. During 
both observations, the XIS detectors were operated in ``burst'' clock 
mode with `1/4' window option yielding 0.5~s time resolution. The 
publicly available archival data (observation~IDs: 905002010 and 
406060010) were used in the present study. Heasoft software package 
of version 6.12 and calibration database (CALDB) for XIS and HXD released
on 2014 February 03 and 2011 September 13, respectively, were used in
the data analysis.  

The unfiltered event data files were reprocessed by applying the `aepipeline'
package of FTOOLS. The clean event files generated after the reprocessing of 
XIS and HXD unfiltered event data were used in our analysis. The arrival times
of the X-ray photons recorded in XIS and HXD event data were corrected for solar 
system barycenter by applying `aebarycen' task of {\small FTOOLS}. The light 
curves and spectra of the pulsar were extracted from clean event data by using 
the $XSELECT$ package of FTOOLS. The XIS event data were corrected for the effect 
of thermal flexing and wobbling by applying attitude correction script 
\textit{aeattcor.sl}\footnote{http://space.mit.edu/ASC/software/suzaku/aeattcor.sl}. 
Subsequently, the pile-up estimation was made for XISs data by using S-lang script 
\textit{pile\_estimate.sl}\footnote{http://space.mit.edu/ASC/software/suzaku/pile\_estimate.sl}.
For first observation, a pile-up of $\sim$22\%, $\sim$18\% and  $\sim$28\%  
was found at the centers of XIS-0, XIS-1 and XIS-3, respectively. Therefore, an 
annulus region with inner and outer radii of $60''$ and $180''$ was selected 
to reduce the effect of pile-up to $\le$4\%. During the second observation, 
the pile-up was estimated to be 17\%, 13\% and 18\% at the centers of XIS-0, 
XIS-1 and XIS-3, respectively. An annular region with $40''$ inner and $180''$ 
outer radii was used to reduce the pile-up effect to $\le$4\% for the second 
observation. These annulus regions were used for the extraction of source
light curves and spectra from cleaned XIS event data. The XIS background light 
curves and spectra were extracted from the event data by selecting a circular 
region away from the source. Response matrices and effective area files for 
XISs were created by using `xisrmfgen' and `xissimarfgen' tasks of FTOOLS, 
respectively. HXD being a non-imaging detector system, source light curves 
and spectra were obtained from cleaned HXD/PIN and HXD/GSO event data by using 
{\it XSELECT} package. However, PIN and GSO background light curves and spectra 
were accumulated from simulated non-X-ray background event files provided by the
instrument team. A correction for cosmic X-ray background 
(CXB\footnote{http://heasarc.nasa.gov/docs/suzaku/analysis/pin\_cxb.html})
was also applied to PIN spectrum. In our spectral analysis, HXD/PIN 
response files released in 2010 July and 2011 June were used for 2010 
August and 2012 January observations, respectively. For GSO data, 
response and effective area files released in 2010 May were used for 
2010 August observations.

\section{Results}

\subsection{Timing Analysis}

As described above, source and background light curves with 1~s time resolution 
were extracted from barycentric corrected XIS-0, PIN and GSO event data for both 
the observations. The $\chi^2$-maximization technique was used to estimate the pulse 
period of the pulsar. Pulsations at periods of 275.45$\pm$0.05 and 274.88$\pm$0.01~s 
were detected in the source light curves obtained from the first and second 
{\it Suzaku} observations, respectively. Quoted errors in pulse period are 
calculated for 90\% confidence level. The estimated pulse periods were used 
to generate pulse profiles from background subtracted light curves from 
corresponding observations. Pulse profiles in 0.5-10~keV (XIS-0), 10-70~keV 
(PIN) and 40-200~keV (GSO) energy ranges for both the observations are shown 
in second, third and fourth panels of Fig.~\ref{lc}, respectively.
Pulse profiles were generated by using 55421.7600 MJD as epoch (phase zero) 
for the first observation where as 55957.4326 MJD was used for the second 
observation.  

Strong energy dependence of pulse profiles can be clearly seen during both
observations (Fig.~\ref{lc}) . Absorption dips at certain phases were seen 
in the soft X-ray pulse profiles (0.5-10~keV range). However, these dips 
disappeared from the pulse profiles in 10-70~keV range. The pulsations were 
absent or marginally seen in 40-200~keV pulse profiles obtained from GSO 
light curves. Apart from the energy dependence, the pulse profiles are also 
found to be luminosity dependent. The top panels of the Fig.~\ref{lc} show 
that the {\it Suzaku} observations of the pulsar were carried out at different 
luminosity levels e.g. the source was comparatively brighter during the 2010 
August observation than the 2012 January observation. However, the shape of 
the pulse profiles in soft and hard X-rays (second and third panels) are 
different due to the presence of absorption dips or dip-like features. Along 
with the energy and luminosity dependence of the pulse profiles in GX~304-1, 
a phase shift of $\sim$0.1 (see Fig.~\ref{lc}) was also found between the soft 
(XIS) and hard (PIN) X-ray  pulse profiles obtained from both the observations.

To investigate the evolution of pulse profiles with energy during both 
{\it Suzaku} observations, we generated energy resolved pulse profiles 
in various energy bands and are shown in Fig.~\ref{erpp2010} \& 
~\ref{erpp2012} for first and second observations, respectively. It 
can be seen from Fig.~\ref{erpp2010} that a prominent and narrow 
absorption dip was present in pulse profiles up to $\sim$8 keV. Beyond 
this energy, the peak in the pulse profiles prior to the dip ($\leq$ 0.3 
phase; left panels of Fig.~\ref{erpp2010}) disappeared and broadened 
the minima in the pulse profile to 0.05-0.25 pulse phase range (panels 
in second column of Fig.~\ref{erpp2010}). We suggest that this broadening 
of the minima in the pulse profile is the possible cause of observed 
$\sim$0.1 phase shift in the soft (0.5-10~keV) and hard X-ray (10-70~keV) 
pulse profiles. A careful inspection of Fig.~\ref{erpp2010} \& 
~\ref{erpp2012} showed that the pulse profiles were complex due to presence
of several absorption features (dips) at various pulse phases and strongly 
energy dependent up to $\sim$12~keV beyond which the shape of the profiles
became simple up to $\sim$35~keV. The main dip in the pulse profiles 
in this energy range (12-35 keV) was observed to be phase-shifted by 
$\sim$0.1 phase compared to that in soft X-ray profiles. In pulse profiles
beyond $\sim$35 keV, the main dip at $\sim$0.1 phase appeared to be filled-up 
gradually with increase in energy. Along with the increase in the normalized
intensity at $\sim$0.1 phase (Fig.~\ref{erpp2010} \& ~\ref{erpp2012}), a 
significant decrease in intensity was observed at $\sim$0.7-0.8 phase range
which appeared as the main dip in hard X-ray pulse profiles. These hard X-ray 
pulse profiles ($\geq$40~keV) appeared to be single-peaked and pulsations were 
detected up to $\sim$120~keV.

\begin{figure}
\centering
\includegraphics[height=3.2in, width=2.4in, angle=-90]{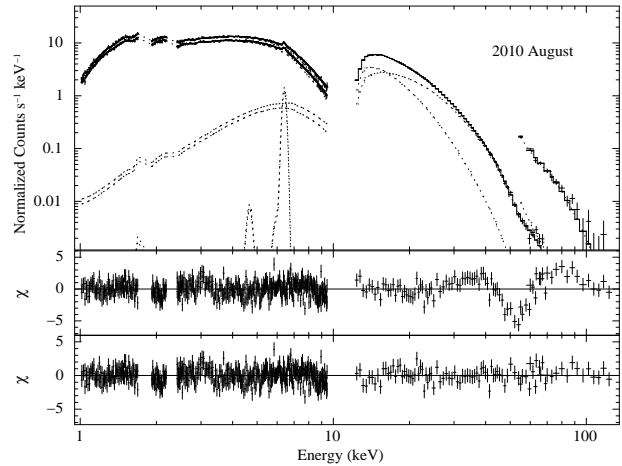}
\caption{Energy spectrum of GX~304-1 in 1-130~keV energy range obtained with the XIS-0, XIS-3, 
PIN and GSO detectors from the first {\it Suzaku} observation in 2010 August outburst, 
along with the best-fit model comprising  a partial covering NPEX continuum model, 
a Gaussian function for iron emission line and a cyclotron absorption component. 
The middle and bottom panels show the contributions of the residuals to $\chi^{2}$ 
for each energy bin for the partial covering NPEX continuum model 
without and with cyclotron component in the model, respectively.}
\label{spec2010}
\end{figure}

\begin{figure}
\centering
\includegraphics[height=3.2in, width=2.4in, angle=-90]{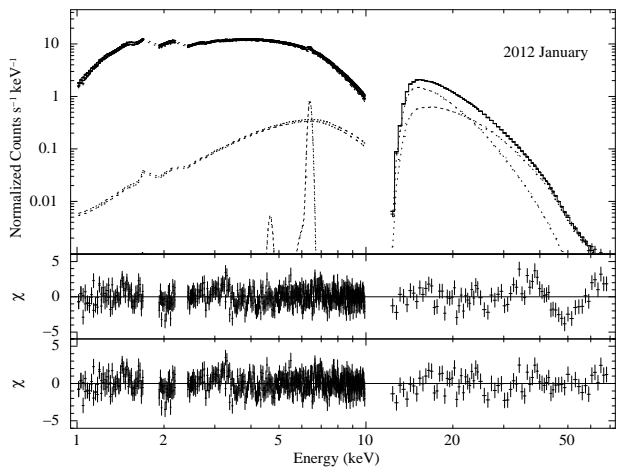}
\caption{Energy spectrum of GX~304-1 in 1-70~keV energy range obtained with the XIS-0, XIS-3 and PIN 
detectors from the second {\it Suzaku} observation in 2012 January outburst, along with the 
best-fit model comprising  a partial covering NPEX continuum model, a Gaussian 
function for iron emission line and a cyclotron absorption component. The middle 
and bottom panels show the contributions of the residuals to $\chi^{2}$ 
for each energy bin for the partial covering NPEX continuum model 
without and with cyclotron component in the model, respectively.}
\label{spec2012}
\end{figure}

The energy resolved pulse profiles of GX~304-1 were found to be complex during
both the {\it Suzaku} observations. Presence of multiple narrow and prominent 
absorption dips were seen up to as high as $\sim$50 keV. Beyond this energy, 
the profiles appeared relatively simple. Presence of prominent dips up to higher 
energies and sudden change in phase of main dip in pulse profiles beyond $\sim$35~keV 
made it interesting to investigate the properties of pulsar through phase-averaged and 
phase-resolved spectroscopy.

\begin{table*}
\centering
\caption{Best-fitting parameters (with 90\% errors) obtained from the spectral fitting of {\it Suzaku} 
observations of GX~304-1 during 2010 August and 2012 January outbursts. 
Model-1 : partial covering NPEX 
model with Gaussian component; Model-2 : partial covering NPEX model 
with Gaussian component and cyclotron absorption line.}
\begin{tabular}{lllll}
\hline
Parameter                      &  \multicolumn{2}{c}{2010 August}     	&   \multicolumn{2}{c}{2012 January}   \\ 
\\
                                &Model-1             &Model-2            &Model-1           &Model-2       \\
\hline
N$_{H1}$$^a$                    &1.04$\pm$0.02       &1.02$\pm$0.02      &0.98$\pm$0.02	    &0.97$\pm$0.02    \\
N$_{H2}$$^b$                    &13.2$\pm$1          &13.7$\pm$1.2       &5.7$\pm$0.5	      &5.3$\pm$0.5    \\
Cov. fraction                   &0.35$\pm$0.02       &0.32$\pm$0.02      &0.25$\pm$0.02	    &0.23$\pm$0.02	   \\
Photon index                    &0.6$\pm$0.03        &0.57$\pm$0.03       &0.42$\pm$0.02	    &0.43$\pm$0.02    \\
E$_{cut}$ (keV)	                &6.6$\pm$0.1         &7.1$\pm$0.2        &6.8$\pm$0.1       &7.4$\pm$0.2    \\

Fe line energy (keV)            &6.41$\pm$0.02       &6.41$\pm$0.02      &6.41$\pm$0.01     &6.41$\pm$0.01 \\
Eq. width of Fe line  (eV)      &44$\pm$7		         &43$\pm$8           &23$\pm$2          &23$\pm$3          \\

Cyclotron line energy (keV)     &--                  &53.2$\pm$0.8       &--                &50$\pm$1      \\
Width of cyclotron line (keV)   &--                  &6.5$^{+2.1}_{-1.6}$ &-- 	            &5.5$^{+2.6}_{-1.8}$   \\
Depth of cyclotron line	        &--                  &0.8$\pm$0.1        &--                &0.5$\pm$0.1 \\

Flux$^c$ (1-10 keV)             &9.2$\pm$0.5        &9.2$\pm$0.6        &4.9$\pm$0.2         &4.9$\pm$0.2 \\
Flux$^c$ (10-70 keV)            &24.5$\pm$1.5       &24.4$\pm$2.0        &9.6$\pm$0.5       &9.6$\pm$1.1   \\ 
Flux$^c$ (70-130 keV)           &0.2$\pm$0.1        &0.2$\pm$0.1         &--                &--           \\
$\chi^2$ (dofs)                 &857 (527)          &638 (524)           &863 (588)		     &765 (585)      \\
\hline
\end{tabular}
\\
\flushleft
$^a$ : Equivalent hydrogen column density in the source direction (in 10$^{22}$ atoms cm$^{-2}$ units),\\ 
$^b$ : Additional hydrogen column density (in 10$^{22}$ atoms cm$^{-2}$ units), \\
$^c$ : Absorption corrected flux in units of 10$^{-9}$  ergs cm$^{-2}$ s$^{-1}$. 
\label{spec_par}
\end{table*}

\subsection{Spectral Analysis}

\subsubsection{Pulse-phase-averaged spectroscopy}

Phase-averaged spectroscopy was performed by using spectra accumulated from XIS-0, 
XIS-1, XIS-3, PIN and GSO data obtained from both the observations. Earlier described 
procedures were followed to obtain source and background spectra, response matrices 
and effective area files for corresponding detectors. After appropriate background 
subtraction, simultaneous spectral fitting was carried out by using {\small XSPEC} 
v12.7 package. As the pulsar was bright during the first observation, broad-band 
spectral fitting was carried out in 1-130~keV range. However, data in 1-70~keV 
energy range were used for simultaneous fitting for second observation. 
The XIS spectra from 2010 August observation were binned by a factor of 4 up 
to 3~keV, a factor of 6 from 3 to 7~keV and a factor of 8 from 7 to 10~keV, whereas
for 2012 January observation, the XIS spectra were binned by a factor of 5 up to 10~keV. 
PIN spectra from both the observations were binned by a factor of 2 from 25~keV to 45 keV, 
a factor of 3 from 45~keV to 50 keV and a factor of 5 from 50 to 70~keV. The GSO spectra 
were grouped as suggested by instrumentation team. Data in 1.7-1.9~keV and 2.2-2.4~keV 
energy ranges were ignored from the spectral fitting due to the presence of known Si and 
Au edge features in the XIS spectra. All the spectral parameters except the relative 
normalization of detectors were tied together during the fitting.

\begin{figure*}
\begin{center}$
\begin{array}{ccc}
\includegraphics[height=5 cm,angle=-90]{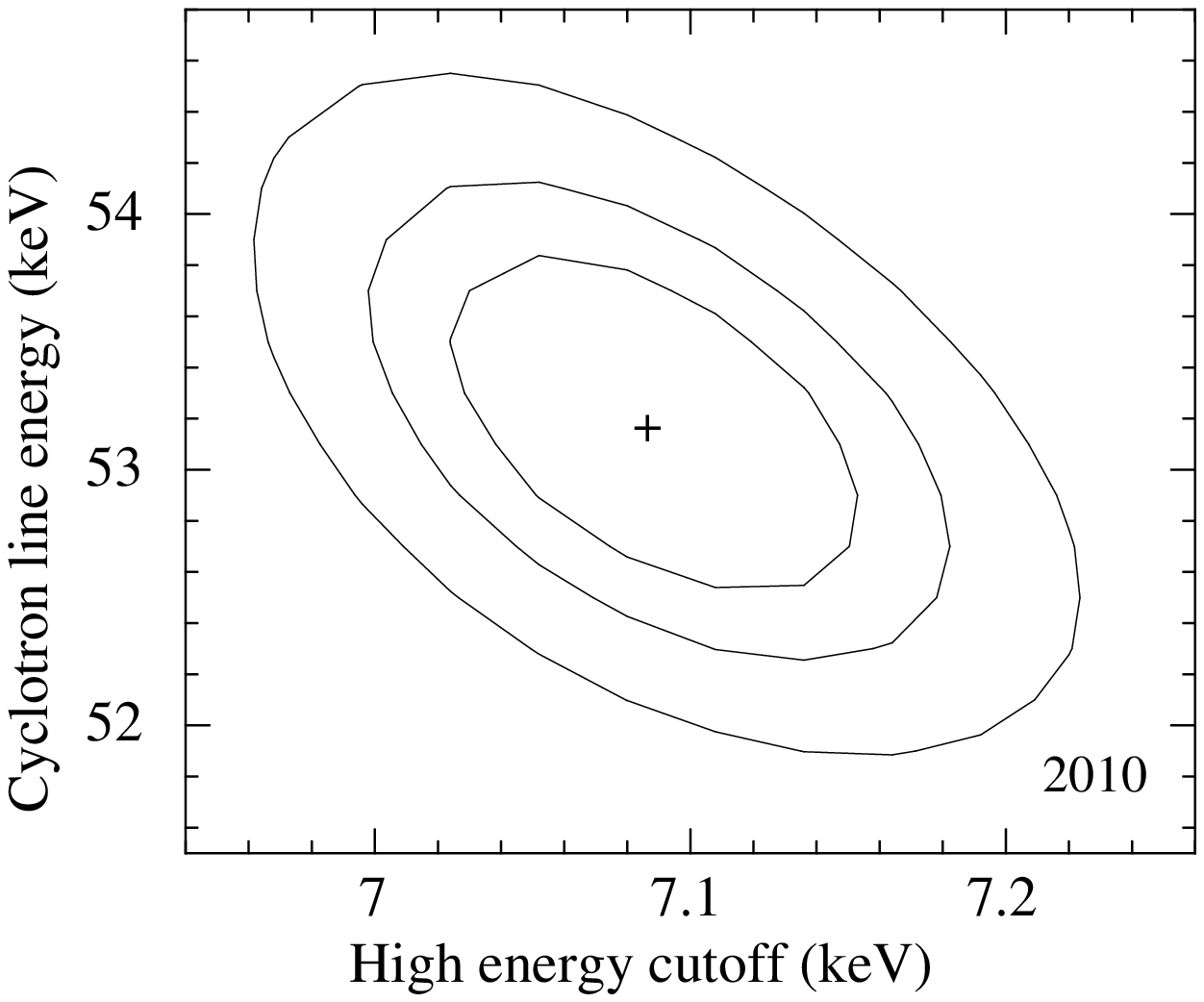} &
\includegraphics[height=5 cm,angle=-90]{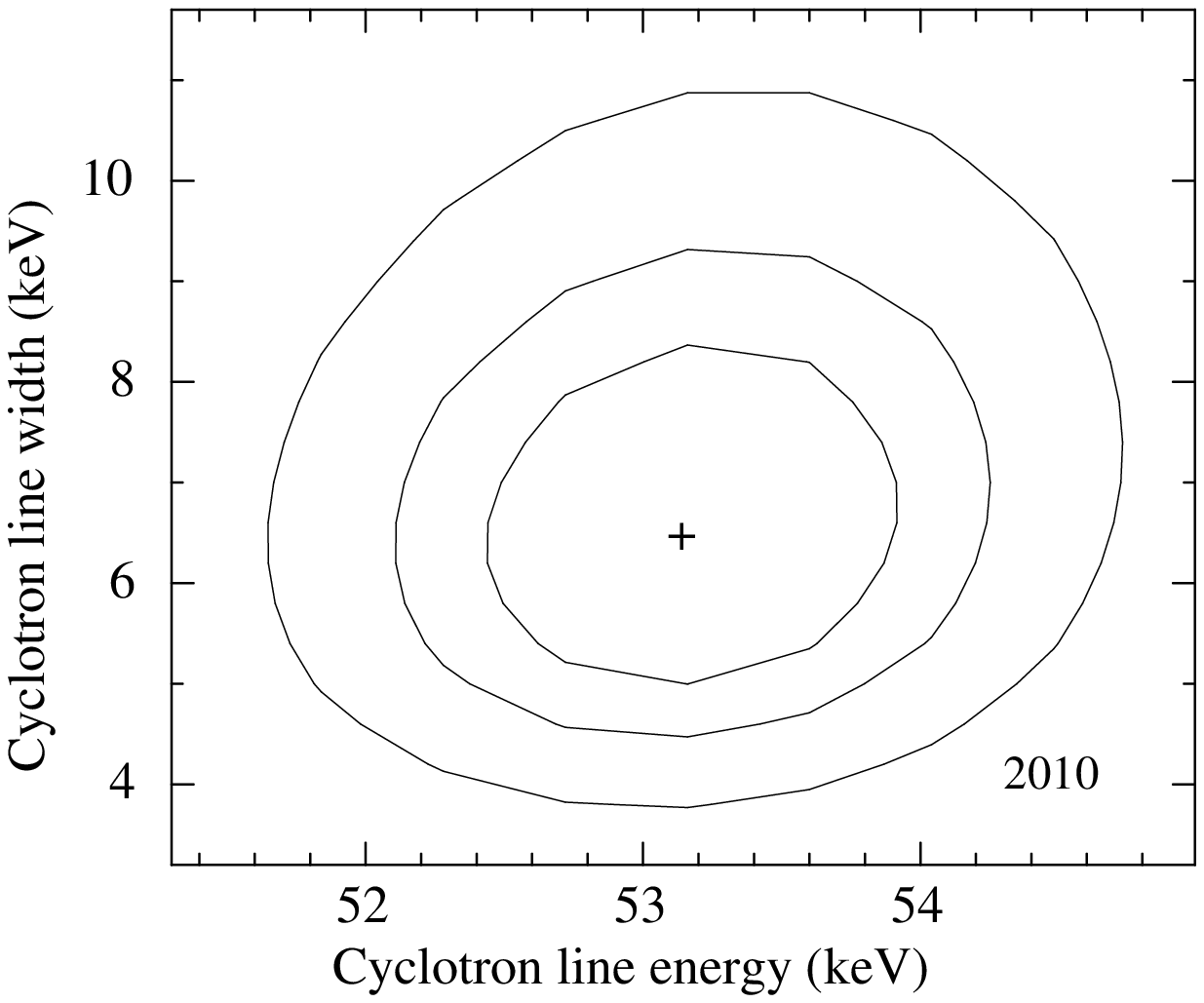} & \\
\includegraphics[height=5 cm,angle=-90]{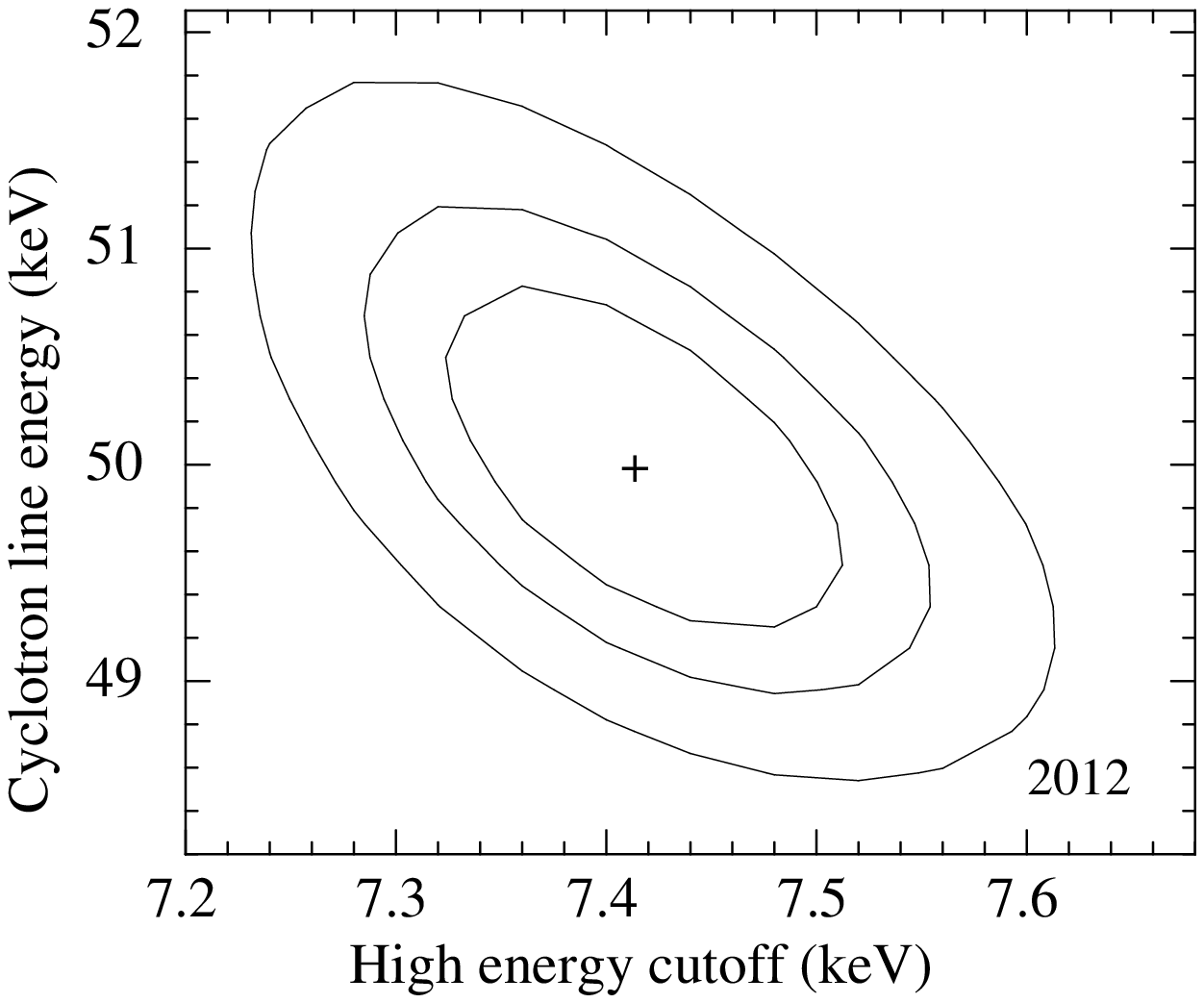} &
\includegraphics[height=5 cm,angle=-90]{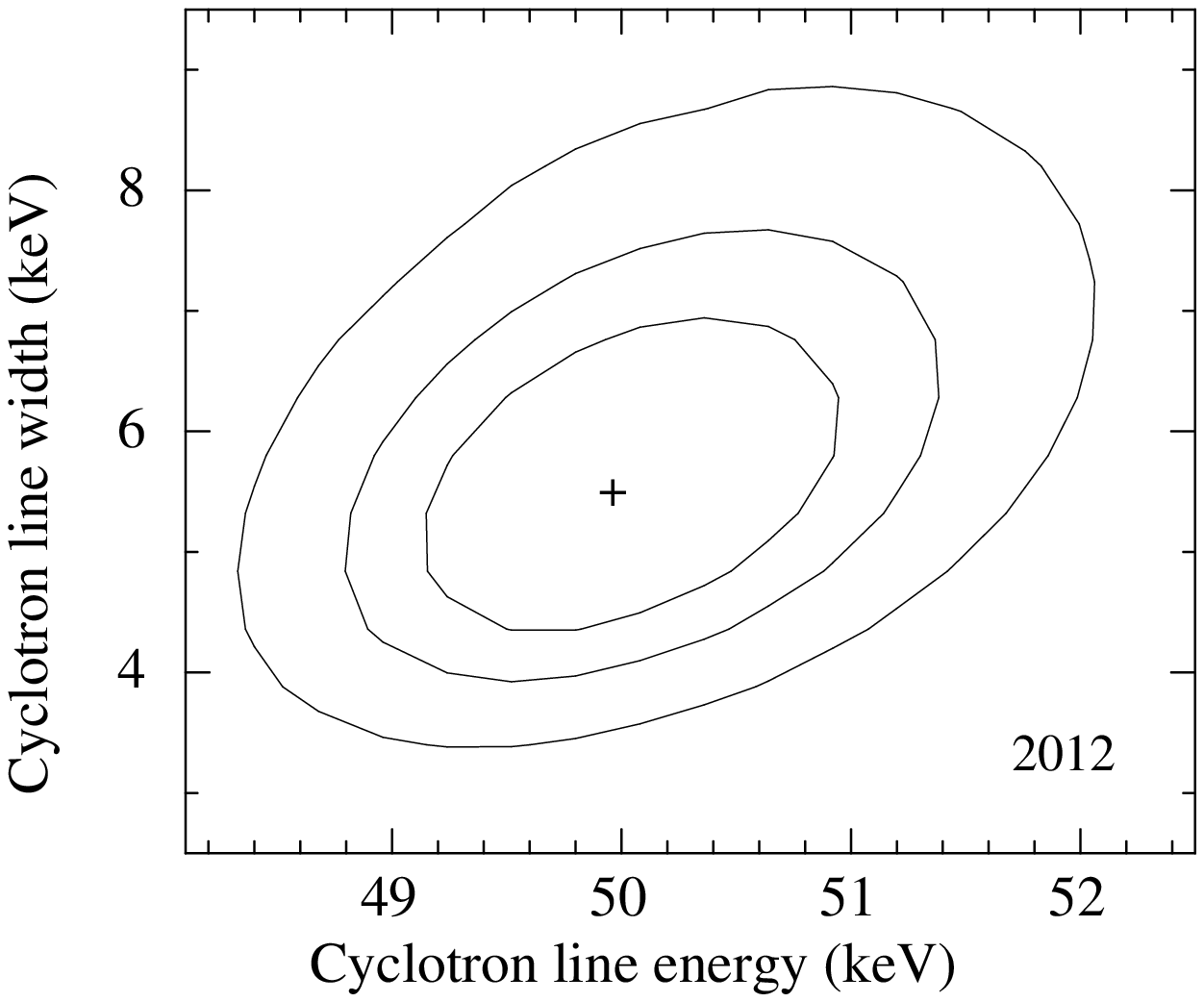} &
 \end{array}$
\end{center}
\caption{The $\chi^{2}$ confidence contours between high energy cutoff, cyclotron line energy and width, 
obtained from the phase averaged spectra fitted by the partial covering NPEX model with cyclotron 
component during 2010 August (upper panel) and 2012 January (lower panel) {\it Suzaku} observations. 
The innermost to outermost contours represent 68$\%$, 90$\%$ and 99$\%$ confidence levels, respectively. 
The ``+" sign indicates the best fit values for both parameters.}
\label{con-all}
\end{figure*}    

It has been found that the phase-averaged spectra of Be/X-ray binary pulsars during 
outbursts have been described by power-law (e.g. GX~304-1; Maurer et al. 1982), high 
energy cutoff power-law (e.g. 4U~0115+63, 4U~1145-61, GX~304-1; White et al. 1983), Fermi
Dirac cutoff power-law (e.g. X0331+53, A~0535+26; Tanaka 1986), NPEX (e.g. 4U~0115+63, 
X0331+53, Cep~X-4; Makishima et al. 1999) continuum models. However, recently it has 
been found that while performing phase-resolved spectroscopy on data taken during X-ray
outbursts of these pulsars, above models do not yield acceptable fit at all phase bins,
specifically at phases of prominent and narrow absorption dips in the pulse profiles
(Naik et al. 2011; Paul \& Naik 2011; Naik et al. 2013 and references therein).
To investigate the changes in spectral parameters at all phase bins (dip and non-dip 
phases of the pulse profiles), a partial covering absorption component has been added
to above standard continuum models. Addition of this component to the continuum model
resulted in getting acceptable fit to all phase bins and explained the cause of the
narrow and prominent dips in the pulse profiles. Partial covering absorption model 
consists of two different power-law components with same photon index but different 
normalizations, being absorbed by different column densities (N$_{H1}$ \& N$_{H2}$), 
respectively (Endo et al. 2000). 

In the beginning, standard continuum models such as high energy cutoff power-law 
(HECUT; White et al. 1983), Fermi Dirac cutoff power-law (FDCUT; Tanaka 1986), 
NewHcut (Burderi et al. 2000), cutoff power-law, NPEX (Makishima et al. 1999) and 
Thermal Comptonization model (CompTT; Titarchuk 1994) were used in our spectral 
fitting to describe the continuum spectrum of GX~304-1. Due to the presence of 
narrow absorption dips in the pulse profiles of the pulsar (previous section) as 
seen in other Be/X-ray binary pulsars, a partial covering absorption component
was added to above continuum models. Among these models, partial covering 
NPEX continuum model was found to fit the source spectra obtained from both the 
observations better than all other continuum models. We selected this model to use 
in phase-averaged and phase-resolved spectral analysis of both the {\it Suzaku} 
observations of GX~304-1.   

The NPEX continuum model is a combination of two power-laws with positive 
and negative indices and a high energy cutoff. This model is an approximation 
of the unsaturated thermal Comptonization in  hot  plasma. The analytical form 
of NPEX model is  

\begin{eqnarray}
\centering
\nonumber 
NPEX(E) = (N_1 E^{-\alpha_1} + N_2 E^{+\alpha_2})  ~ \exp \left( -\frac{E}{kT} \right)
\label{eq1}
\end{eqnarray} 

where $N_1$, $\alpha_1$ and $N_2$, $\alpha_2$ are the normalization and photon index 
of the negative and positive power laws, respectively. $kT$ represents the cutoff energy 
in the unit of keV. The photon index of positive power law is fixed at a value of 2, 
representing Wien's peak. During the spectral modeling, we found that the partial 
covering NPEX model described the 1-130~keV (2010 observation) and 1-70~keV (2012
observation) spectra of the pulsar well. In addition to the continuum, iron fluorescence
emission line at $\sim$6.4~keV was detected in spectrum during both the observations. A 
weak iron emission edge like feature was found in the residual during the fitting of 
second observation. This was modeled by the addition of an edge component at $\sim$7.7~keV 
in the spectral model. 

An absorption like feature at $\sim$54~keV was clearly seen in residuals obtained 
from spectral fittings of both the observations. Addition of a cyclotron absorption 
component (`CYCLABS' in {\it XSPEC} package) in the partial covering NPEX 
continuum model improved spectral fitting further with reduced $\chi^2$ of 
$\sim$1.5 for both the observations. The values of reduced $\chi^2$, though acceptable,
are found to be large. Investigating the residuals in 1-10 keV range, we noticed a marginal 
cross calibration uncertainties present between XIS-1 (back illuminated CCD) and XIS-0 \& 3
(front-illuminated CCD) spectra. While fitting the broad-band spectra without considering 
data from XIS-1, the values of reduced $\chi^2$ obtained are 1.22 and 1.31 for first
and second observations, respectively. Best-fit model parameters obtained from 
simultaneous spectral fitting of XIS-0 \& 3, PIN and GSO data are given in Table~1. 
The energy spectra of the pulsar along with the best-fit model components are shown 
in Fig.~\ref{spec2010} and ~\ref{spec2012}. The middle and bottom panels in each 
figure show the residuals to the best-fit model without and with the addition of 
cyclotron absorption feature in the continuum model, respectively. The $\chi^{2}$ 
confidence contours were plotted to check the dependence of the cyclotron line energy 
on high energy cutoff and cyclotron width and are shown in the top and bottom panels 
of Fig.~\ref{con-all} for first and second observations, respectively. We did not 
find any strong degeneracy among these spectral parameters.

\begin{figure*}
 \centering
 \includegraphics[height=4.8in, width=3.9in, angle=-90]{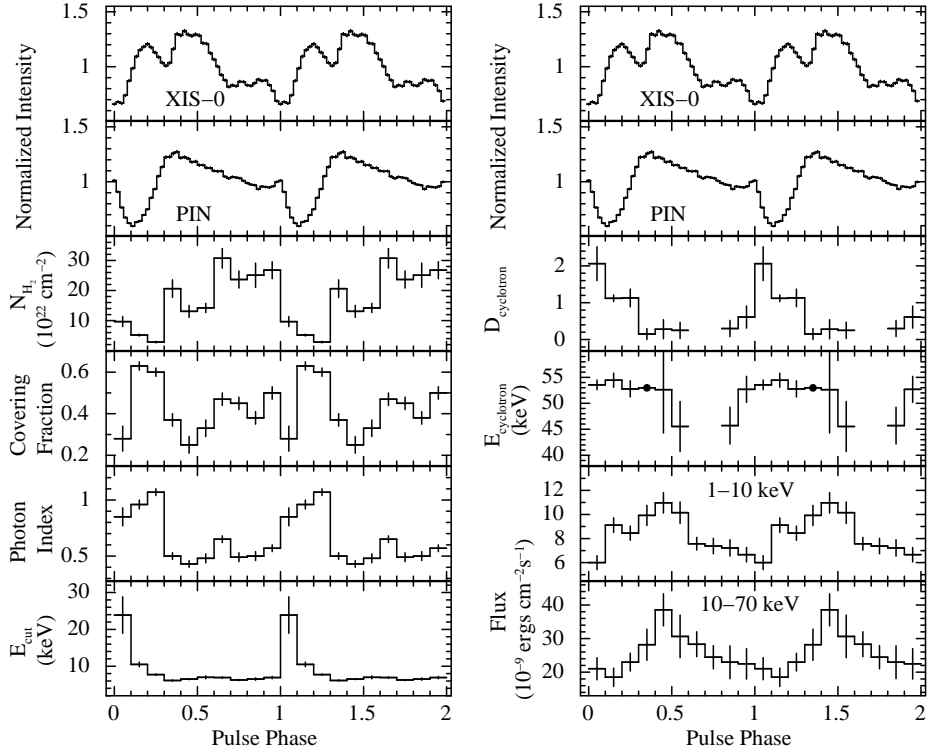}
 \caption{Spectral parameters obtained from the phase-resolved spectroscopy 
of GX~304-1 during the ${\it Suzaku}$ observation in 2010 August.
The first and second panels in both sides show pulse profiles of the pulsar 
in 0.5-10~keV (XIS-0) and 10-70~keV (HXD/PIN) energy ranges. The values of 
$N_{H_2}$, covering fraction, power-law photon index and cutoff energy 
($E_{cut}$) are shown in third, fourth, fifth and sixth panels in left 
side, respectively. The cyclotron line parameters such as depth (third panel),
line energy (fourth panel), source flux in 1-10 keV (fifth panel) and 10-70 keV 
(sixth panel) are shown in right side of the figure. Solid circles in the fourth
panel in right side indicate that the cyclotron line energy was fixed for 
corresponding phase-bin at the phase-averaged value. The errors in the figure 
are estimated for 90\% confidence level.}
\label{sp2010}
 \end{figure*}

\begin{figure*}
\centering
\includegraphics[height=4.8in, width=3.9in, angle=-90]{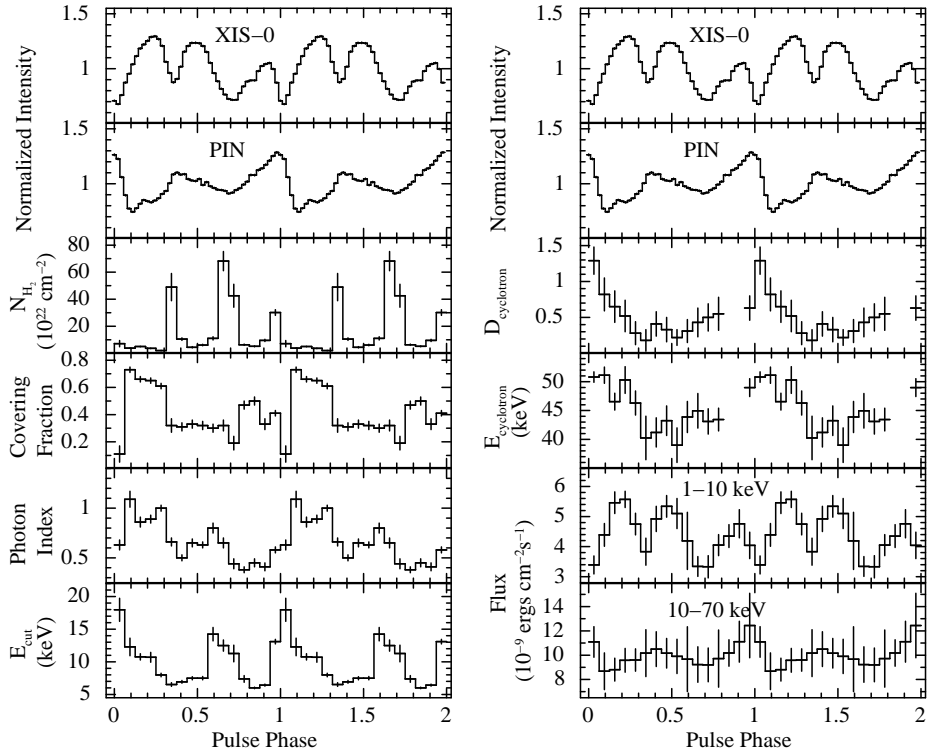}
\caption{Spectral parameters obtained from the phase-resolved spectroscopy 
of GX~304-1 during the ${\it Suzaku}$ observation in 2012 January.
The first and second panels in both sides show pulse profiles of the pulsar 
in 0.5-10~keV (XIS-0) and 10-70~keV (HXD/PIN) energy ranges. The values of 
$N_{H_2}$, covering fraction, photon index and cutoff energy ($E_{cut}$) 
are shown in third, fourth, fifth and sixth panels in left side, 
respectively. The cyclotron line parameters such as depth (third panel),
line energy (fourth panel), source flux in 1-10 keV (fifth panel) and 10-70 keV 
(sixth panel) are shown in right side of the figure. The errors in the figure 
are estimated for 90\% confidence level.}
\label{sp2012}
 \end{figure*}

\subsubsection{Pulse-phase-resolved spectroscopy}

Pulse profiles of GX~304-1 are found to be different even when compared 
at same energy range from both {\it Suzaku} observations. Therefore, 
it is interesting to perform phase-resolved spectroscopy to probe the changes 
in spectral parameters with pulse phase and then compare 
with the observations at different luminosity level. As the first 
observation was for a relatively short exposure ($\sim$13 ks for HXD) 
compared to the second observation ($\sim$59 ks for HXD), phase-resolved 
spectroscopy was carried out by accumulating source spectra in 10 and 16 
pulse-phase bins for first and second observations, respectively. 
Reprocessed XIS and PIN event data from both the observations were used
to extract phase-resolved spectra by applying phase filter in $XSELECT$ package. 
Data from HXD/GSO detector were not used in the phase resolved spectroscopy 
due to lack of sufficient hard X-ray photons in each phase bin. Background
spectra and response matrices used in phase-averaged spectroscopy were also
used in phase-resolved spectral fitting. Simultaneous spectral fitting was
carried out on phase-resolved spectra obtained from both the observations
by using partial covering NPEX continuum model along with the cyclotron
absorption component and Gaussian function for the iron emission line.
While fitting, the values of relative instrument normalizations, equivalent 
hydrogen column density (N$_{H_1}$) and iron emission line parameters were 
fixed at the corresponding values obtained from phase-averaged spectroscopy. 
The width of the cyclotron absorption line was fixed at the 
phase averaged value to constrain the feature. Spectral parameters obtained
from fitting the XIS and PIN phase resolved spectra are shown in 
Fig.~\ref{sp2010} and ~\ref{sp2012} for 2010 August 13 and 2012 
January 31 observations, respectively. Pulse profiles obtained from 
XIS-0 and PIN event data of each observation are also shown in top 
two panels of these figures. 

During both the observations, all the spectral parameters showed significant 
variability with pulse-phase of the pulsar. Fig.~\ref{sp2010} (2010 August 
observation) shows that the value of additional hydrogen column density 
(N$_{H_2}$) varies in 1-30 $\times$ 10$^{22}$~cm$^{-2}$ range over pulse 
phases. The value of N$_{H_2}$ was increased from 0.3 phase and became
maximum in 0.6-1.0 phase range. It can be seen that normalized intensity (XIS 
pulse profile; top panel) was low at 0.6 phase and remain steady before reaching 
the minimum value at 1.0 phase. Decrease in the normalized intensity of the pulsar 
during above pulse phase range and simultaneous increase in the value of N$_{H_2}$
confirm that the plateau like feature in XIS pulse profile in 0.6-1.0 pulse phase
range was due to the presence of additional absorbing material close to the pulsar.
This can also explain the presence of an absorption dip in the XIS pulse profile 
and an increase in the N$_{H_2}$ value in 0.3-0.4 phase range of the pulsar. As 
in case of N$_{H_2}$, the covering fraction of the additional absorption was also 
found to be comparatively high during the dip and plateau region. As expected, 
the values of power-law photon index and cutoff energy were high 
during the main dip in the XIS and PIN pulse profiles. The depth of the 
cyclotron absorption feature was found to be variable and single peaked. 
The energy of cyclotron absorption feature also showed variation over the 
pulse phase of the pulsar. The values of both the parameters were maximum at 
dip phase of the pulsar and gradually decreased to minimum value in 0.5-0.8 
phase range. Data gaps in third and fourth panels in second column are due 
to the non-detection of cyclotron features in the spectral fitting for 
corresponding phase-bins. The flux in 1-10~keV and 10-70~keV shows  
variation with pulse phase and follows the shape of pulse profile in respective 
energy bands.

Though the shape of XIS and PIN pulse profiles of the pulsar were significantly
different during second {\it Suzaku} observation, the variation of spectral parameters
over pulse phase was comparable. High values of absorption column density
were found at phases of absorption dip in XIS pulse profile. The values of power-law 
photon index, cutoff energy, depth and energy of cyclotron absorption line, source 
flux in 1-10 keV and 10-70 keV were followed similar pattern as seen during
the 2010 August {\it Suzaku} observation of the pulsar. Though the number of absorption 
dips and shape of the pulse profiles of the pulsar during both the observations were
different, presence of additional matter as the cause of absorption dips in the soft
X-ray profiles is supported by the findings in the present work. However, low value
($\le$5$\times$10$^{22}$ cm$^{-2}$) of absorption column density at $\sim$0.1-0.2 phase 
range (Fig.~\ref{erpp2010} \& ~\ref{erpp2012}) suggested that the cause of absorption 
dip at above phase range in the hard X-ray pulse profiles was different.

\section{Discussion and Conclusions}

Pulse profiles of transient X-ray binary pulsars are complex due to 
the presence of multiple absorption dips/features at lower energies. 
These absorption features are strongly energy dependent 
-- prominent in soft X-ray pulse profiles and gradually disappear at 
higher energies. Complex structures or absorption dips in pulse profiles 
originate due to the photo-electric absorption of soft X-ray 
photons by matter present around the neutron star. This can be confirmed 
by investigating energy resolved pulse profiles and the evolution of spectral 
properties with pulse phases of the pulsar. Among the accretion powered 
X-ray pulsars, the pulse profiles of Be/X-ray binary pulsars during outbursts 
are found to be complex. These pulsars show regular and periodic X-ray outbursts 
(Type~I) that are associated with the periastron passage of the neutron star. 
During periastron passage, the neutron star captures copious amount of matter 
from the circumstellar disk of the Be companion star and undergoes X-ray 
outbursts. Typical X-ray luminosity of the pulsar during Type~I outburst 
is  $\sim$10$^{36-37}$~erg~s$^{-1}$ (Negueruela et al. 1998). In case of 
GX~304-1, Type~I X-ray outbursts occur at a period of 132.5~d. Using  
instruments with good time resolution and wide-band energy coverage 
capabilities onboard {\it Suzaku}, we carried out a detailed study of
the pulsar during two Type~I outbursts. We also performed pulse-phase 
resolved spectroscopy of GX~304-1 to study the evolution of spectral 
parameters during these outbursts.

\subsection{Pulse Profiles}

Pulse profiles of GX~304-1 appeared to be different
during 2010 August and 2012 January Type~I outbursts. 
During 2010 August observation, the profiles were complex due to the
presence/absence of peaks and dips at several pulse phases in soft and 
hard X-ray energy ranges. Due to this, a phase shift of $\sim$0.1 was visible 
between the soft and hard X-ray pulse profiles (Fig.~\ref{lc} \& ~\ref{erpp2010}). 
However, the shape of the pulse profiles of the pulsar during 2012 January outburst 
were significantly different compared to 2010 August outburst. Number of absorption
dips in the pulse profiles was more during 2012 observation. The strength of 
these dips was also prominent compared to the earlier observation. These dips 
were present in the pulse profiles up to higher energies unlike the 2010 August 
observation, where the dips were seen in the pulse profiles up to $\sim$10 keV. 
Pulse-phase resolved spectroscopy during both the observations revealed the 
presence of additional matter at certain phases, causing the absorption dips 
in the soft X-ray pulse profiles. During 2012 observation, more number of 
narrow absorption dips in the pulse profiles suggest the presence of dense 
narrow streams of matter around the neutron star. However, during 2010 observation, 
relatively simpler profile with single absorption dip and a plateau phase suggests 
that the matter distribution around the neutron star is different and relatively 
simple. The luminosity of the pulsar during 2010 {\it Suzaku} observation was 
estimated to be higher (2.3$\times$10$^{37}$ergs~s$^{-1}$) than the 2012 
{\it Suzaku} observation (1$\times$10$^{37}$ergs~s$^{-1}$). The peak luminosity 
of 2010 August outburst was also found to be high compare to the 2012 January 
outburst (first panels of Fig.~1). As both the outbursts are Type~I X-ray outbursts,  
observed difference in the shape of the pulse profiles must be due to the difference 
in the subsequent mass accretion rate. 

As in case of GX~304-1, another transient Be/X-ray binary pulsar EXO~2030+375
was also observed at the peaks of 2007 May and 2012 May Type-I outbursts with 
{\it Suzaku}. The pulse profiles of EXO~2030+375 during these two outbursts were 
significantly different (Naik et al. 2013; Naik \& Jaisawal 2015). During 
the luminous outburst in 2007 May, the pulse profile of EXO~2030+375 consisted 
of several absorption dips as seen in the pulse profiles of GX~304-1 during 2012 
January outburst. However, during the less intense 2012 May outburst, the pulse 
profiles were relatively smooth. Difference in the mass accretion rate during 
these two Type~I outbursts was interpreted as the cause of different shape of 
pulse profiles in EXO~2030+375. Such type of pulse profiles with multiple dips 
are also seen in other Be/X-ray binary pulsars such as A0535+35 (Naik et al. 2008), 
GRO~J1008-57 (Naik et al. 2011), 1A~1118--61 (Maitra et al. 2012 and references 
therein). The dips in pulse profiles of these pulsars are originated due to the 
absorption of emitted radiation by matter close to the neutron star. In most of 
the cases, the dips in the pulse profiles are seen only in soft X-rays. However, 
there are a few Be/X-ray pulsar in which the absorption dips are seen at same 
phase in the pulse profiles up to higher energy ($\sim$70~keV) e.g. EXO~2030+375 
(Naik et al. 2013). This is interpreted as due to the presence of dense additional 
absorbers at various narrow pulse phase bins of the pulsar. Though, dip-like features 
in GX~304-1 (present work) appeared up to $\sim$50~keV, the low value of additional 
column density at dip phases suggests that absorption is not the cause of these 
dips in hard X-ray pulse profiles.

Energy resolved pulse profiles of GX~304-1, obtained from both the {\it Suzaku}
observations revealed a significant phase-shift ($\sim$0.35) of the main dip in 
profiles at energies below and above $\sim$35~keV. The observed phase-shift of
the main dip happened to occur at an energy close to the cyclotron absorption 
line energy in GX~304-1. Such type of effects e.g. phase-shifts (lags) or 
significant variations in pulse profiles close to the cyclotron absorption 
line energy are also seen in two other Be/X-ray binary pulsars such as V~0332+53 
(Tsygankov et al. 2006) and 4U~0115+63 (Ferrigno et al. 2011). Using a numerical
study on the effect of cyclotron resonant scattering in highly magnetized accretion 
powered X-ray pulsars, Sch{\"o}nherr et al. (2014) showed that a strong change in 
the pulse profile is expected at the cyclotron absorption line energy. This change
is attributed to the effects of angular redistribution of X-ray photons by cyclotron 
resonant scattering in a strong magnetic field combined with relativistic effects.
In GX~304-1, we observed a significant change in the phase of main dip in the pulse
profile close to the cyclotron line energy. This detection, along with
the reported results from V~0332+53 and 4U~0115+63 supports the results obtained from
the numerical study of Sch{\"o}nherr et al. (2014).

\subsection{Spectroscopy}

Broad-band spectra of accretion powered X-ray pulsars are described by 
several continuum models such as HIGHECUT, FDCUT, cutoff power-law, NPEX, 
CompTT. Along with the continuum model, additional components such as 
absorption due to matter present in the interstellar medium, blackbody/bremsstrahlung 
for soft X-ray excess, Gaussian functions for emission lines and cyclotron resonance 
scattering features (CRSF) are also needed to explain the observed spectrum. In 
case of several Be/X-ray binary pulsars, a partially absorption component is being 
used to describe the presence of several absorption features (dips) at certain phases 
in the pulse profiles (Paul \& Naik 2011 and references therein). Detection of 
CRSF in the broad-band pulsar spectrum provides direct estimation of
the magnetic field of X-ray pulsars. However, studies of CRSF at different pulse phases 
of the pulsars can reveal important information about the magnetic field geometry around 
the neutron star. Therefore, broad-band spectroscopy of data obtained from the 
observations with high spectral capability instruments on-board {\it Suzaku} 
is an appropriate tool to understand the properties of accretion powered X-ray pulsars.

This paper reports the broad-band phase-averaged and phase-resolved 
spectroscopy of GX~304-1 by using two {\it Suzaku} observations during 
its Type~I outbursts. The estimated values of galactic equivalent hydrogen 
column density in the source direction (N$_{H_1}$) was same (within errors) 
during both the observations. However, the values of additional column density 
(N$_{H_2}$ - local to the pulsar), were found to be different during both 
the observations. The value of N$_{H_2}$ was high during the 2010 August 
observation compared to 2012 January observation. It should be noted here 
that the luminosities of the pulsar during the {\it Suzaku} observation 
(Table~1) as well as at the peak (top panels of Fig.~1) of 2010 August 
outburst were high compared to corresponding values during 2012 January 
outburst. This confirms that the amount of mass accreted by the neutron 
star from the Be circumstellar disk during 2010 August outburst was 
significantly more than that during 2012 January outburst. Similar findings 
have also been reported for EXO~2030+275 during its Type~I outbursts with 
significantly different luminosities (Naik et al. 2013; Naik \& Jaisawal 
2015). High value of N$_{H_2}$ during more luminous Type~I outbursts in 
Be/X-ray binary pulsars can be explained as due to significant amount of 
mass accreted by the neutron star from the circumstellar disk of the Be 
companion star.

We estimated the effect of additional column density (N$_{H_2}$) on the phase 
averaged spectrum by taking the difference in absorption corrected flux without and 
with N$_{H_2}$ component. It was found that the 2010 August observation was more 
affected by absorption due to additional matter compared to the 2012 January 
observation. The flux differences were estimated to be (1.8$\pm$0.9$)\times$10$^{-9}$ 
and (0.4$\pm$0.3$)\times$10$^{-9}$ ergs~cm$^{-2}$~s$^{-1}$ for 2010 August and 2012 
January observations, respectively. The covering fraction which is defined as 
{\it Norm2/(Norm1+Norm2)}, where $Norm1$ and $Norm2$ are the normalizations of 
first and second power-law, was also found to be high during first observation. 
Covering fraction describes the possibility of interaction of photons with 
the absorbing matter which is marginally offset from the line of sight at a phase 
averaged solid angle of {\it $\Omega$/4$\pi$= Norm2/Norm1} (Endo et al. 2000). The 
averaged phase solid angles ($\Omega$/4$\pi$) for first and second observations were 
estimated to be 0.47 and 0.30, respectively. This indicates that the additional 
absorbing matter during first observation was present relatively more closer 
to the neutron star that extending to a larger solid angle. 

Absorption dips in the pulse profiles of Be/X-ray binary pulsars are explained
as due to the presence of narrow streams of matter that are phase-locked with the 
neutron star. The magnetospheric radius of the pulsar can be used to constrain the 
location of additional absorbing matter in such narrow streams. The size of 
magnetosphere of the pulsar depends on the mass accretion rate and the strength 
of the magnetic field (Mushtukov et al. 2015a and references therein) through the 
relation  
\begin{equation}
R_{\rm m}=2.6\times 10^8
 M^{1/7}R_6^{10/7}B_{12}^{4/7}L_{37}^{-2/7} \ \mbox{cm}
\end{equation}  
Using standard values of mass and radius of canonical neutron stars and the observed 
luminosities of GX~304-1 in above equation, the magnetospheric radii were estimated
to be $\sim$5100 and 6300~km during first and second observations, respectively. 
This confirmed that the pulsar magnetosphere was relatively small during the first
observation. Therefore, the streams of additional matter were present within the 
magnetosphere and found to be closer to the pulsar during the first {\it Suzaku} 
observation.

Pulse-phase resolved spectroscopy of GX~304-1 during two Type~I outbursts 
showed the presence of narrow streams of matter (higher value of N$_{H_2}$) at 
several pulse phase ranges. The presence of such streams of matter are interpreted 
as the cause of absorption dips detected in soft X-ray pulse profiles of the pulsar. 
Several such dips are seen in the pulse profiles of Be/X-ray binary pulsars during 
X-ray outbursts. Density and opacity of matter in these narrow streams decides the 
energy dependence of the dips in the pulse profiles of these pulsars. During both 
the observations, pulse-phase variation of covering fraction was found to be similar. 
We suggest that the geometry or distribution of absorbing matter around the pulsar 
is probably similar during both observations but at different locations. 

\subsection{Cyclotron line in GX~304-1}

Cyclotron absorption line or CRSF was clearly detected in the
spectra of GX~304-1 during both the observations and was described 
with pseudo-Lorentzian function (CYCLABS) in the continuum model. 
These are absorption like features which appear in hard X-ray spectrum 
($\sim$10-100~keV) through the resonant scattering of the photons with 
electrons in strong magnetic field ($\sim$10$^{12}$~G) near the poles of 
the neutron star. The cyclotron line energy is related with the magnetic 
field through the relation {\it E$_{cyc}$=11.6B$_{12}\times(1+z){^{-1}}$ 
(keV)} or 12-B-12 rule, where B$_{12}$ is the magnetic field in the unit 
of 10$^{12}$~G and $z$ is the gravitational red-shift. Detection of 
CRSF in the pulsar spectrum, therefore, provides  the direct measurement
of the magnetic field of the neutron star. Though, CRSF was already detected 
in GX~304-1 (Yamamoto et al. 2011), {\it Suzaku} observations 
at different source intensities provide an opportunity to 
investigate the luminosity dependence of the cyclotron line energy. 
The cyclotron feature was detected at $\sim$53 and 50~keV during 2010 August 
and 2012 January {\it Suzaku} observations, respectively. The observed 
cyclotron absorption line energies are not significantly different.  
However, it shows a positive dependence on the pulsar luminosity, as seen 
in Her~X-1 (Staubert et al. 2007). In some other cases such as 4U~0115+63 
(Nakajima et al. 2006; Tsygankov et al. 2007), and V~0332+53 (Tsygankov et al. 
2010), a negative correlation is observed between the luminosity of pulsar 
and the cyclotron absorption line energy.

Attempts have been made to explain the observed positive and negative 
correlation between pulsar luminosity and the cyclotron line 
energy. A negative correlation is expected in super-critical regime 
where the source luminosity is higher than the critical luminosity 
(Becker et al. 2012; Mushtukov et al. 2015a). Above critical luminosity, 
density of infalling matter becomes so high that the particles start 
interacting and decelerating through the formation of a radiation-pressure 
dominated shock above the neutron star surface. Cyclotron scattering features 
most likely occur closer to the shock region. With increasing luminosity, the 
shock height drift upwards in the accretion column where relatively low 
strength of magnetic field produces cyclotron feature at lower energy. It 
explains negative correlation as observed in 4U~0115+63 and V~0332+53. However, 
the positive correlation is predicated for a sub-critical regime e.g. below 
critical luminosity. The pressure of accreting matter in this regime pushes 
hydrodynamical shock closer to the neutron star surface where increase in 
the strength of magnetic field with luminosity results a positive correlation 
between the cyclotron line energy and luminosity (Staubert et al. 2007), as seen 
in Her~X-1 and GX~304-1. The 1-70 keV luminosity of GX~304-1 during first and 
second {\it Suzaku} observations was estimated to be 2.3$\times$10$^{37}$ 
and 1$\times$10$^{37}$ergs~s$^{-1}$, respectively, which lies in the sub-critical 
regime (Mushtukov et al. 2015a) and expected to show a positive correlation  
 with cyclotron line energy.

Poutanen et al. (2013) and Nishimura (2014) have proposed alternate models 
to describe the dependence of cyclotron absorption line energy on the source
luminosity. In these models, it is argued that the formation of cyclotron line 
in direct spectrum is questionable due to the large gradient of magnetic field 
strength along line forming region in the accretion column. However, cyclotron 
line (CRSF) can be formed in the reflected radiation from the surface of the 
neutron star where magnetic field gradient is relatively small (Poutanen et al. 
2013). It is widely believed that the accretion column height is linearly 
dependent on source luminosity ($>$10$^{37}$~ergs~s$^{-1}$) or accretion rate. 
At higher luminosity, larger area on neutron star surface (from the poles with 
strong magnetic field towards equator with weak field strength) is expected to 
be illuminated for cyclotron interactions. This explains the observed 
anti-correlation between cyclotron line energy and luminosity, as seen in 
V~0332+53. However, Nishimura (2014) described the observed correlations by 
considering the changes in polar emission region, direction of photon 
propagation and the shock height. The negative correlation was interpreted 
in terms of shock region displacement whereas the positive correlation was 
explained by using the changes in beam pattern. In case of GX~304-1 (present
work), the cyclotron absorption line width, depth and ratio of line width and 
energy (W/E$_{cyl}$) show positive correlation with luminosity. The observed
positive correlation between the cyclotron line energy and luminosity in 
GX~304-1, therefore, can be explained as due to the change in beam pattern,
as described by Nishimura (2014). Alternatively, Mushtukov et al. (2015b) 
discussed the positive luminosity dependence of sub-critical X-ray pulsars 
like GX~304-1 by studying the changes of plasma velocity profile in line-forming 
region under influence of radiation pressure from the hot-spot. The cyclotron line 
energy at a luminosity was determined by corresponding redshift from
velocity profile at a given height. This model successfully predicts 
the positive luminosity correlation of cyclotron line energy for pulsars 
like GX~304-1 along with cyclotron line width dependence on luminosity and
pulse phase variation of cyclotron line parameters, which were not being 
discussed in Nishimura (2014).    

Investigation of the change in cyclotron parameters with pulse phase 
of the pulsar provides important evidences to understand the emission 
geometry as well as the magnetic field mapping around the neutron star. 
A comparative study of these parameters at different intensity levels 
during several outbursts can also yield information about the changes 
in emission or accretion column geometry. Pulse-phase resolved spectroscopy 
of cyclotron absorption line is one of the interesting results of this work. 
During both {\it Suzaku} observations of the pulsar, cyclotron energy and depth 
were found significantly variable with pulse-phase. Change in cyclotron 
line parameters with pulse phase has also been seen in other X-ray pulsars such 
as Cen~X-3 (Burderi et al. 2000), GX~301-2 (Kreykenbohm et al. 2004), 1A~1118-61 
(Maitra et al. 2012), A~0535+26, XTE~J1946+274, 4U~1907+09 (Maitra \& Paul 2013), 
V~0332+53 (Lutovinov et al. 2015) and Cep~X-4 (Jaisawal \& Naik 2015) with 
$\sim$10-30\%  variation in cyclotron energy. 

The cyclotron line energy for GX~304-1 was found variable with phase up 
to 17\% and 24\% for first and second {\it Suzaku} observations, respectively. 
However, the cyclotron line depth showed phase dependence within a factor of 
$\sim$2.5 for both observations. Various simulations were performed to study the 
dependence of the cyclotron line on pulsar phases, assuming certain sets of 
rule and geometry in line forming region (Sch{\"o}nherr et al. 2007; Mukherjee 
\& Bhattacharya 2012). These studies predicated that 10-30\% variation  
in cyclotron energy with phases can be resulted due to effects of viewing angle 
of accretion column or emission region. However, more than 30\% changes in 
cyclotron energy is contributed by distortion in the magnetic field geometry.
In GX~304-1 (present work), cyclotron energy is found to be variable within 
24\% for both observations which can be explained as effects of viewing angle 
or local distortion in the magnetic field in line forming region. We found that 
the cyclotron energy and depth for both observations were peaking around the 1.0-1.1 
(or 0.0-0.1) phase range, which lies closer to the peak of pulse profiles at 
$\geq$35~keV. This result supports the effect of cyclotron resonance 
scattering on beaming pattern that changes the shape of the pulse profiles.    

In summary, the timing and spectral properties of GX~304-1 were presented by using 
{\it Suzaku} observations during Type~I outbursts. Absorption dips were detected in the 
pulse profiles at soft X-rays. These dips were originated due to obscuration/absorption 
of X-ray photons with narrow streams of accretion flow at certain pulsar phases.
The shape of pulse profiles was significantly changed at $\geq$35~keV e.g. close
to the cyclotron line energy. Significant pulse phase dependence of cyclotron line 
was explained as the effects of viewing angle or the role of complicated magnetic 
 field of the pulsar.

\section*{Acknowledgments}

We sincerely thank the referee for his/her valuable comments and suggestions which 
improved the paper significantly. The research work at Physical Research Laboratory 
is funded by the Department of Space, Government of India. The authors would like to 
thank all the members of the Suzaku for their contributions in the instrument 
preparation, spacecraft operation, software development, and in-orbit instrumental 
calibration. This research has made use of data obtained through HEASARC Online 
Service, provided by the NASA/GSFC, in support of NASA High Energy Astrophysics 
Programs.


\begin{thebibliography}{}


\bibitem[]{}Becker P., Klochkov D., Sch{\"o}nherr G., et al. 2012, A\&A, 544, A123 
\bibitem[]{}Bradt H. V., Clark G. W., Dower R., Doxsey R., Hearn D. R., Jernigan J. G., et al. 1977, Nature, 269, 21
\bibitem[]{}Burderi L., Di Salvo T., Robba N. R., La Barbera A., Guainazzi M., 2000, ApJ, 530, 429
\bibitem[]{}Corbet R. H. D., Smale A. P., Menzies J. W., Branduardi-Raymont G., Charles P. A., 
            et al., 1986, MNRAS, 221, 961
\bibitem[]{}Devasia J., James M., Paul B., Indulekha K., 2011, MNRAS, 417, 348
\bibitem[]{}Endo T., Nagase F. and Mihara T., 2000, PASJ, 52, 223
\bibitem[]{}Ferrigno C., Falanga M., Bozzo E., Becker P. A., Klochkov D., Santangelo A., 2011, A\&A, 532, A76
\bibitem[]{}Forman W., Jones C., Cominsky L., Julien P., Murray S., et al. 1978, ApJS, 38, 357
\bibitem[]{}Giacconi R., Murray S., Gursky H., Kellogg E., Schreier E., et al. 1974, ApJS, 27, 37
\bibitem[]{}Haefner R., 1988, Inf. Bull. Var. Stars, 3260, 1
\bibitem[]{}Jaisawal G. K. and Naik S., 2015, MNRAS, 453, L21
\bibitem[]{}Klochkov D., Doroshenko V., Santangelo A., Staubert R., Ferrigno C., et al. 2012, A\&A, 542, L28
\bibitem[]{}Koyama K. et al., 2007, PASJ, 59, 23
\bibitem[]{}Kreykenbohm I., Wilms J., Coburn W. et al., 2004, A\&A, 427, 975 
\bibitem[]{}Krimm H. A.,  et al., 2010, Astron. Telegram, 2538, 1
\bibitem[]{}Lutovinov A. A., Tsygankov S. S., Suleimanov V. F. et al., 2015, 448, 2175
\bibitem[]{}Maitra C., Paul B., Naik S., 2012, MNRAS, 420, 2307
\bibitem[]{}Maitra C. \& Paul B., 2013, 771, 96
\bibitem[]{}Makishima K., Mihara T., Nagase F., Tanaka Y., 1999, ApJ, 525, 978
\bibitem[]{}Manousakis A., et al. 2008, Astron. Telegram, 1613
\bibitem[]{}Mason K. O., Murdin P. G., Parkes G. E., Visvanathan N., 1978, MNRAS, 184, 45P
\bibitem[]{}Maurer G. S., Johnson W. N., Kurfess J. D., Strickman M. S., 1982, ApJ, 254, 271
\bibitem[]{}Malacaria C., Klochkov D., Santangelo A., \& Staubert R., 2015, A\&A, 581, 121
\bibitem[]{}McClintock J. E., Ricker G. R., Lewin W. H. G., 1971, ApJ, 166, L73
\bibitem[]{}McClintock J. E., Rappaport S. A., Nugent J. J., Li F. K.,  1977, ApJ, 216, L15
\bibitem[]{}Mihara T., et al.  2010, Astron. Telegram, 2779
\bibitem[]{}Mitsuda K. et al., 2007, PASJ, 59, 1
\bibitem[]{}Mukherjee D., \& Bhattacharya D., 2012, MNRAS, 420, 720
\bibitem[]{}Mushtukov A. A., Suleimanov V. F., Tsygankov S. S. and Poutanen J., 2015a, MNRAS, 447, 1847
\bibitem[]{}Mushtukov A. A., Tsygankov S. S., Serber A. V., Suleimanov V. F., Poutanen J., 2015b, MNRAS, 454, 2714
\bibitem[]{}Naik S., Dotani T., Terada Y., et al. 2008, ApJ, 672, 516
\bibitem[]{}Naik S., Paul B., Kachhara C., Vadawale S. V., 2011, MNRAS, 413, 241
\bibitem[]{}Naik S., Maitra C., Jaisawal G. K., Paul B., 2013, ApJ, 764, 158
\bibitem[]{}Naik S. \& Jaisawal G. K., 2015, RAA, 15, 537
\bibitem[]{}Nakajima M., Mihara T., Makishima K., Niko H.,  2006, ApJ, 646, 1125
\bibitem[]{}Negueruela I., Reig P., Coe M. J., \& Fabregat J., 1998, A\&A, 336, 251
\bibitem[]{}Nishimura O, 2014, ApJ, 781, 30
\bibitem[]{}Parkes G. E., Murdin P. G., Mason K. O., 1980, MNRAS, 190, 537
\bibitem[]{}Paul B. \& Naik S., 2011, Bull. Astron. Soc. India, 39, 429 
\bibitem[]{}Pietsch W., Oegelman H., Kahabka P., Collmar W., Gottwald M., 1986, A\&A, 163, 93
\bibitem[]{}Priedhorsky W. C., Terrell J., 1983, ApJ, 273, 709
\bibitem[]{}Postnov K. A., Mironov A. I., Lutovinov A. A., Shakura N. I., Kochetkova A. Yu., Tsygankov S. S., et al. 2015, MNRAS, 446, 1013
\bibitem[]{}Poutanen J. et al., 2013, ApJ, 777, 115
\bibitem[]{}Sch{\"o}nherr G., et al. 2007, A\&A, 472, 353
\bibitem[]{}Sch{\"o}nherr G., et al. 2014, A\&A, 564, 8
\bibitem[]{}Staubert R., Shakura N. I., Postnov K., et al. 2007, A\&A, 465, L25
\bibitem[]{}Sugizaki M., Yamamoto T., Mihara T., Nakajima M., \& Makishima K., 2015, PASJ, 67, 73
\bibitem[]{}Takahashi T. et al., 2007, PASJ, 59, 35  
\bibitem[]{}Tanaka Y., 1986, in Proc. IAU Colloq. 89, Vol. 255, Radiation
            Hydrodynamics in Stars and Compact Objects, ed. D. Mihalas \& K.-H. A. Winkler  
\bibitem[]{}Titarchuk L., 1994, ApJ, 434, 313 
\bibitem[]{}Tsygankov S. S., Lutovinov A. A., Churazov E. M., \& Sunyaev R. A., 2006, MNRAS, 371, 19
\bibitem[]{}Tsygankov S. S., Lutovinov A. A., Churazov E. M., \& Sunyaev R. A., 2007, Astron. Lett., 33, 368
\bibitem[]{}Tsygankov S. S., Lutovinov A. A., \& Serber A. V., 2010, MNRAS, 401, 1628
\bibitem[]{}Yamamoto T.  et al., 2009, Astron. Telegram, 2297, 1
\bibitem[]{}Yamamoto T., Sugizaki M., Mihara T., Nakajima M., Yamaoka K., et al. 2011, PASJ, 63, 751
\bibitem[]{}Yamamoto T.,  et al., 2012, Astron. Telegram, 3856, 1
\bibitem[]{}White N. E., Swank J. H.  \& Holt S. S., 1983, ApJ, 270, 711

\end{thebibliography}
\end{document}